# Computational Validation of a Mathematical Model of Stable Multi-Species Communities in a Hawk Dove Game

**Jeffrey Lu**
York Community High School
jeffreylu2021@u.northwestern.edu

*We revisit the original hawk-dove game with slight modifications to payoff values while maintaining the fundamental principles of interaction. The practical robustness of the theoretical tools of game theory is tested on a simulated population of hawks and doves with varying initial population distributions and peak growth rates. Additionally, we aim to find conditions in which the entire community fails or becomes a single-species population. The results show that the predicted community distribution is established by the majority of communities but fails to exist in communities with extreme initial imbalances in species distribution and insufficient growth rates. We also find that greater growth rates can compensate for more imbalanced initial conditions and that more balanced initial conditions can compensate for lower growth rates. Overall, the simple theoretical model is a strong predictor of the stable behavior of simulated multi-species communities.*[1]

## 1    Introduction

Some situations, take Rufai et al. (2021) for example, that apply mixed strategy equilibrium closely resemble the classic prisoners' dilemma game (Axelrod 1987), in which 2 players have two strategy options with no strategy being absolutely superior to the other. One common "real-world" example of this is the hawk-dove game (Smith 1982), which presents a biological analog to the prisoners' dilemma. The optimal mixed-strategy equilibrium can be easily calculated with a linear system based on the payoff matrix. The result of this calculation can be interpreted as a prediction of the distributions of the 2 strategies within a stable community. Within a biological model, however, there are always additional environmental factors that impact population dynamics and, as a result, may impact the population distributions within the community.

In this paper, the mixed-strategy equilibrium predictions of population distributions are validated through computational simulations of an isolated community with two competing species of birds representing the 2 strategies. The results indicate that, while there are certain extreme cases

---

in which a multi-species community fails to be established, the predictions from a trivial mixed-strategy equilibrium model very accurately reflect the population distributions in the simulated stable multi-species communities.

## 2 Methods

### 2.1 Simulation Configuration

We elect to use a modified version of the simple hawk-dove game (Eldakar 2020), which was originally proposed by Smith (1982). Our hawk-dove game involves competition between 2 species of birds, hawks and doves, for food and, as a result, survival. Every day, the birds go out to eat and randomly pair up with another bird to distribute food. In this model, "hawks" are aggressive and will fight the other bird for food if they do not get the majority. "Doves" are passive and will split the food equally if the other bird allows it but will yield to hawks. When hawks are paired with each other, they continuously fight and therefore receive no utility. When a hawk is paired with a dove, the hawk happily takes the majority of the food and the dove is left with little. The hawk receives maximum utility and the dove receives limited utility. When doves are paired with each other, they split the food and both receive a high utility.

|  |  | Player 2 | |
|---|---|---|---|
|  |  | **Hawk** | **Dove** |
| Player 1 | **Hawk** | 0, 0 | $1, \frac{1}{4}$ |
|  | **Dove** | $\frac{1}{4}, 1$ | $\frac{11}{12}, \frac{11}{12}$ |

Table 1: The payoff matrix of the hawk-dove game used in the simulations. Each value is a pair of survival probabilities with the left and right elements corresponding to the survival probabilities of players 1 and 2 respectively.

Each population is allowed to independently grow exponentially at the same maximum rate, though this is naturally limited by a finite amount of food that "naturally" appears each day. Because birds pair up randomly for food, the probability that any given species will have members that do not receive food is directly proportional to the size of the population.

All simulations used an initial population size of 1000 birds with a theoretical maximum carrying capacity of 1000 birds. Both the initial distribution of hawks and doves and the maximum growth rate varied across experiments. Each simulation is run for 200 days with the population counts recorded at the end of each day.

## 2.2 Theoretical Predictions

The mixed-strategy equilibrium for the payoff matrix given in Table 1 can be calculated using a linear system of equations.

Let $h \in [0, 1]$ be the proportion of the population that is hawks and $d \in [0, 1]$ be the proportion of the population that is doves. We know $h + d = 1$ because the theoretical population is composed entirely of hawks and doves.

The equilibrium is found at the strategy distribution for one player which ensures the payoffs for the other player are the same for all possible strategies. Thus, $0h + 1d = \frac{1}{4}h + \frac{11}{12}d$.

A system of the previous 2 equations can be solved to yield $h = 0.25$ and $d = 0.75$. Essentially, the payoffs alone predict that the communities will converge to a distribution of 25% hawks and 75% doves. It follows that the survival probabilities for both hawks and doves are 75% when equilibrium is reached. Therefore, we can arrive at the prediction that, given enough time, all communities will have a stable dove population fluctuating around 562.5 and a stable hawk population fluctuating around 187.5.

## 3 Results and Discussion

Simulations testing general convergence patterns were run in sets of 100. We find that the vast majority of populations within randomly varied initial population distributions and maximum growth rates converge to and maintain the theoretical predictions.

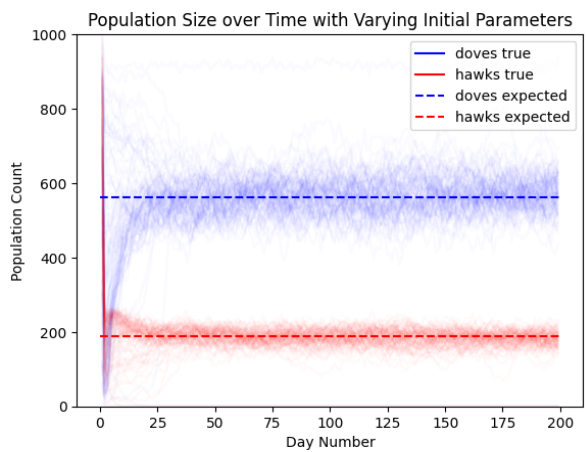

Figure 1: The raw population counts of hawks and doves over 100 200-day simulations with randomly varying dove proportions and maximum growth rates.

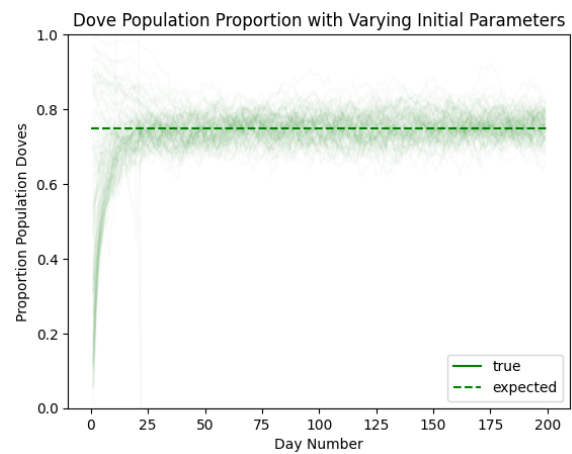

Figure 2: The ratio of dove population to total population over 100 200-day simulations with randomly varying dove proportions and maximum growth rates

## 3.1 Conditions for Failed Communities

In certain circumstances, the community becomes completely occupied by doves or becomes void of birds altogether. These circumstances seemed to occur in low initial dove proportions or low maximum growth rates.

To investigate these potential trends, we explored the *stable population sizes* at varying values of initial dove proportions and maximum growth rates. The stable population size for a given simulation is determined by the arithmetic mean of the population sizes from the last 100 days.

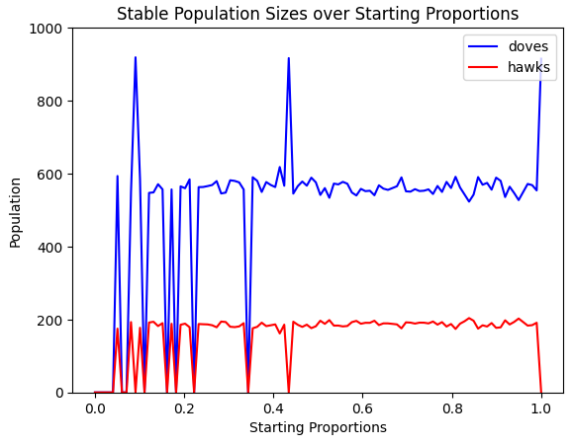

Figure 3: The stable population sizes of hawks and doves over 100 200-day simulations with initial dove proportions between 0% doves and 100% doves with a constant maximum growth rate of 0.5.

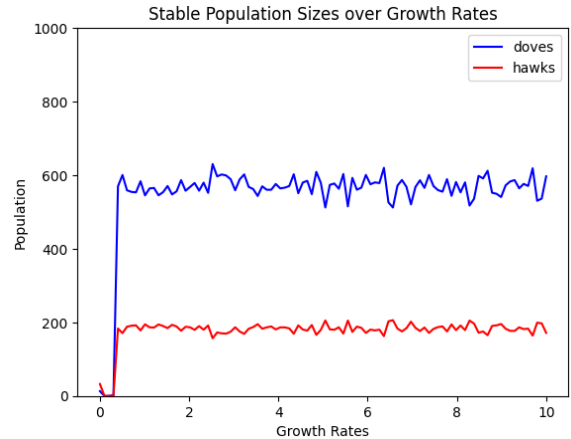

Figure 4: The stable population sizes of hawks and doves over 100 200-day simulations with maximum growth rates between 0 and 10 with a constant initial dove proportion of 0.75.

In both Figure 3 and Figure 4, stable multi-species communities do not exist for values very close to 0. Communities that fail to establish stable population sizes of at least 100 for both hawks and doves, which we will refer to as *failed multi-species communities*, have 2 main variants. The more common variant involves the death of almost all birds in the community. This occurs in both low initial dove proportion and low maximum growth rate. Another variant of failed multi-species communities involves the extinction of hawks but not doves. Because a population of doves results in a non-zero survival probability, a sufficient growth rate can keep it alive and stable. Note that this cannot occur with a hawk-only population because hawks paired with hawks result in 0% survival rate (Table 1).

It follows that there likely exists some relationship between both population parameters that impact multi-species community dynamics differently than each parameter does individually. We measured this relationship using the average rate of community establishment, which is the rate at which a given set of initial parameters produces a non-failed multi-species community. We simulated 2700 input parameter combinations 30 times to approximate the relationship between initial dove proportion, maximum growth rate, and community establishment rate. We validated the preliminary finding that lower values of both parameters decrease community establishment.

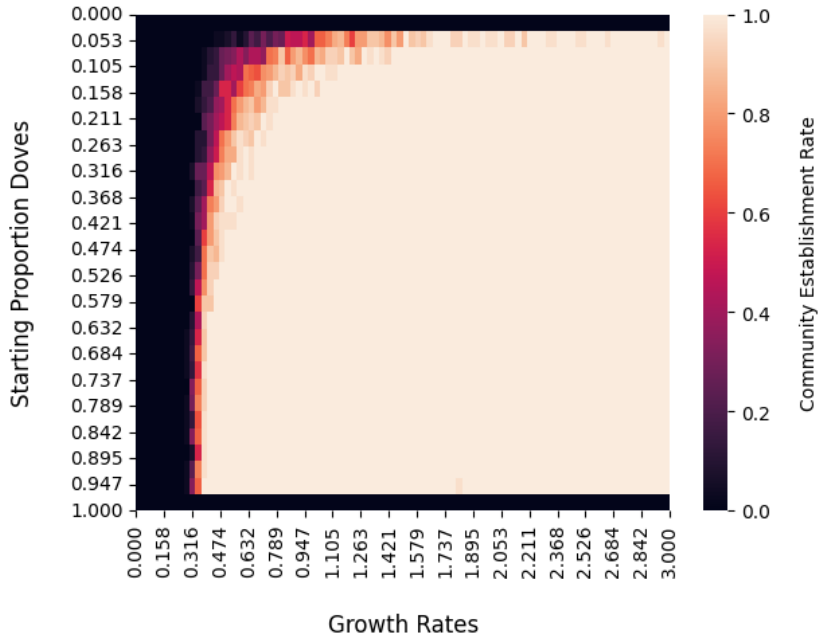

Figure 5: The average community establishment rates over growth rates between 0 and 3 and starting proportion doves between 0 and 1.

Interpreting Figure 5, there are clear trends that show high community establishment at growth rates beyond a certain point of sufficiency and non-extreme values of initial dove population proportions. Additionally, there is a clear relationship between growth rates and initial proportions, namely that higher growth rates can compensate for lower initial dove proportions and higher initial dove proportions can compensate for lower growth rates.

## 4    Conclusion and Future Work

In this paper, we evaluated the strength of a simple mathematical model for a variation of the classic game theory example of hawks and doves using simulations. We found that the simulated behavior closely mirrors the predicted behavior in most situations, though low population growth rates and extreme imbalances in initial population distribution consistently cause divergence from the predicted values. Though trivial, simple models have been shown to provide useful information for predicting the behavior of larger and more complex systems (May 1971).

Further investigation may explore the potential to predict the conditions for stable multi-species community establishment using theoretical tools. This work was also limited in some aspects, such as only investigating the effect of 2 community parameters on stability and only predicting stable population properties instead of the path to stability. Additional exploration to examine additional parameters, as well as properties of unstable populations, is needed.

Though there are still many areas that require more work, we found that, while intuitive and commonplace, simple game theory models can still be powerful predictors for the behavior of real-world phenomena.

## 5    Acknowledgements

This paper was originally written for *Game Theory, and Practice* in 2021 at Northwestern University and was later revised to add analysis of the conditions for failed communities and improve formatting. Special thanks to Professor Scott Ogawa for assisting with project ideation and initial review. This work was also supported by Deepnote's generous computing resources for research.